\documentclass[onecolumn,showpacs,preprintnumbers]{revtex4}
\usepackage[english]{babel}
\usepackage{graphicx}
\usepackage{dcolumn}
\usepackage{ifthen}
\usepackage{bm}

\begin{document}
%\preprint{APS}

\title{Black hole state evolution and Hawking radiation }

\author{ D. Ahn{\footnote{E-mail:dahn@uos.uos.ac.kr; daveahn@ymail.com }} \thanks{Also with the Institute of Quantum Information Processing and Systems, University of Seoul, Seoul 130-743, Republic of Korea} \thanks{E-mail; dahn@uos.uos.ac.kr }
        }

\address{Department of Electrical and Computer Engineering, University of Seoul, Seoul 130-743, Republic of Korea \\
        }

\date{\today}

%\maketitle
%\wideabs{
\begin{abstract}
The effect of a black hole state evolution on the Hawking radiation is studied using the final state boundary condition. It is found that thermodynamic or statistical mechanical properties of a black hole depend strongly on the unitary evolution operator $S$ which determines the black hole state evolution. When the operator $S~$ is random unitary or pseudo random unitary, a black hole emits thermal radiation as predicted by Hawking three decades ago. On the other hand, it is found that the emission of Hawking radiation could be suppressed when the evolution of a black hole state is given by the generator of the coherent state.
\end{abstract}

\pacs{PACS numbers: 03.67.-a, 89.70.+c}

\maketitle

%}
\pagebreak
%\noindent{\bf Introduction}

For over three decades, the Hawking radiation has been mainly the subject of black hole information paradox \cite{Hawk3} until recently the possibility of microscopic black hole production by the Large Hadron Collider (LHC) became manifested \cite{Dimopoulos,Giddings1}. Hawking's semiclassical argument predicts that a process of black hole formation and evaporation is not unitary \cite{Hawk3}. On the other hand, there is evidence from string theory that the formation and evaporation of black holes should be consistent with the basic principles of quantum mechanics \cite{HM}.

Possible microscopic black hole production by the LHC would be a very exciting manifestation of the fundamental physics   \cite{Dimopoulos,Giddings1}.
These microscopic black holes are expected to undergo the prompt and quasi-thermal evaporation by emitting Hawking radiation  \cite{Hawk1, Hawk2}.
There have been extensive studies on the modeling of black hole evaporation
  \cite{Fabbi,Dai,Konoplya} and whether a meta-stable black hole production in the LHC would be possible or not \cite{Unruh1,Vilkovisky, Giddings2, Casadio, Koch}. It is known that there are examples of absolutely stable black holes, when such black holes represent the ground state of some object such as a topological charge \cite{Mazur}. Then, it would be an interesting problem to study under which condition a meta-stable black hole can be created.

In this study, we investigate the effect of black hole internal quantum state evolution on the Hawking radiation using the final state boundary condition \cite{HM, Ahn1, Ahn2, Ahn3} and show that the emission of Hawking particles would be suppressed for black holes with mass in the TeV range when the black hole internal matter state is in the coherent state. The description of astrophysical object by a coherent state was originally considered for the study of wormholes \cite{Coleman, Preskill, Hawk4}. It is later extended to the case of black hole problem \cite{Fiola}. The coherent state is an eigenstate of the annihilation operator $a$ \cite{Barnett}. The coherent states saturate minimum uncertainty bound and hence are good candidates for semi-classical states \cite{Barnett}. It is suggested that the formation of a black hole with internal state described by a coherent state is possible when the short distance feature of the particle $r$  is much smaller than the impact parameter $b$ in high energy particle collisions \cite{Hsu}. We recover Hawking's original results when the internal evolution is governed by the pseudo random unitary operator. Our results demonstrate that the black hole evaporation is strongly dependent on the black hole internal quantum states.

Recently, the author \cite{Ahn1,Ahn2,Ahn3} studied a final wave function for the interior of a Schwarzschild black hole and found that the detailed structure of a black hole internal state is given by a two-mode squeezed state consisting of collapsing matter and infalling Hawking radiation. The concept of the final wave function for the interior of the black hole is closely connected with the idea that the wave function of the universe is unique \cite{Hartle}.

If there is a unique quantum state associated with the big bang singularity in the Planck scale, then there should be a unique quantum state associated with the black hole singularity \cite{HM}. If a matter on a 3-brane collapses under the gravity to form a black hole, then the metric can be approximated by the Schwarzschild metric \cite{Dadhich}.

\begin{figure}[htbp]

 \centering
 \includegraphics[width=0.65\textwidth,angle=-90]{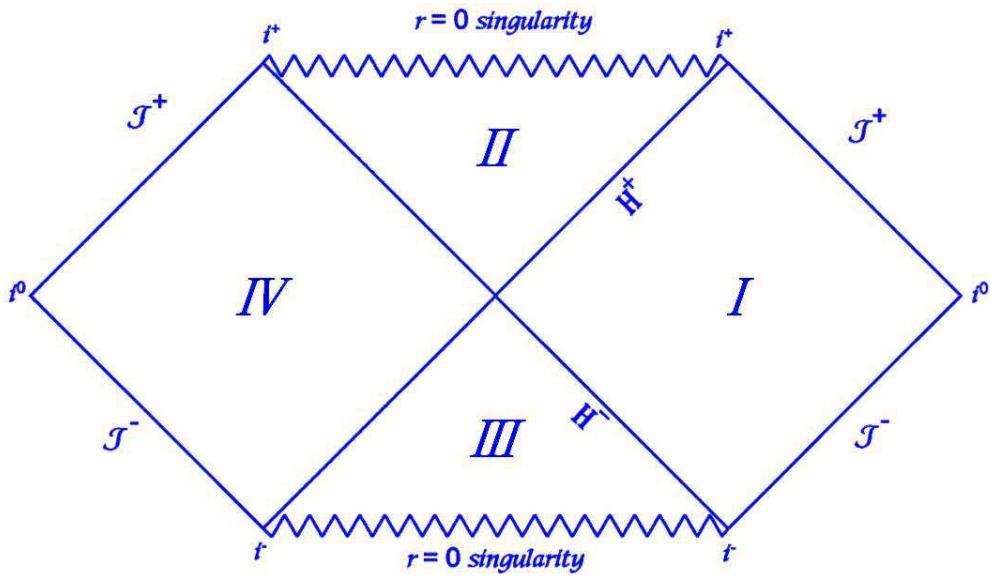}
%  \includegraphics[scale=0.5]{fig2.eps}
%   \special[scale=0.50]{pict=fig_1.pict}
  \caption{The Kruskal extension of the Schwarzschild spacetime \citep{Wald, Davies}.
In region $I$, null asymptotes $H_{+}$ and $H_{-}$ act as future and past horizons, respectively. The boundary lines labeled  $J^{+}$ and $J^{-}$  are future and past null infinity respectively, and $i^{0}$ is the spacelike infinity.}
%  \label{fig:1}
\end{figure}

The final wave function is related to time symmetric formulation of quantum mechanics \cite{Aharanov}. In this formulation, the conditional probability of the system to experience a chain of projection (or measurements) denoted by the operator $C(\alpha)$ when both initial and final states are fixed, is given by
%\end{document}
\begin{equation}
P(\tilde{c}(\alpha)|\tilde{a},\tilde{b})=\frac{Tr[\rho_{b} C(\alpha) \rho_{a} C^{\dagger}(\alpha)]}{\sum_{\alpha} Tr[\rho_{b} C(\alpha) \rho_{a}C^{\dagger}(\alpha)]},
\end{equation}
%\end{document}
where $\rho_{a}=|a \rangle \langle a|~$ is the initial state at the past null infinity $J^{-}$,
$\rho_{b}=|b \rangle \langle b|~$ is the state at the future null infinity $J^{+}$
and $\alpha=(\alpha_{1},...,\alpha_{n})~$ at $(t_{1},...,t_{n})~$ is a sequence of alternatives. Here, $\tilde{a}, \tilde{b}, \tilde{c}(\alpha)~$ are eigenvalues or sequence of eigenvalues.

Let $\{|n\rangle \}$ be the basis of infinite
dimensional Hilbert space $H$.
We assume that the initial matter state which collapses to a black hole belongs to $H_{M}~$. The inital state
at the past null infinity $J^{-}$ (Fig. 1) is assumed to be the
tensor product of the inital matter state in $H_{M}~$
and the Unruh vacuum
state \cite{Unruh2, Wald, Davies} belonging to $H_{in}\otimes H_{out}$.
On the other hand, the final state at the future null infinity $J^{+}$
is assumed to be the product state of an outgoing wave in $H_{out}~$ and a modified
Unruh vacuum state belonging to $H_{M}\otimes H_{in}$. The author has shown
that the ground state of a collapsing shell inside the Schwarzschild black hole
can be obtained by the Bogoliubov transformation of the pair of creation and annihilation
operators belonging to $H_{M}~$ and $H_{in}~$ \citep{Ahn1}.
%\end{document}

%2010.10.19

We
consider only the chain of projections $C_{out}(\alpha)~$ associated with
an outside observer and assume that $\rho_{a}=\rho_{M} \otimes \rho_{in \otimes out}~$ and $\rho_{b}=\tilde{\rho}_{M \otimes in} \otimes \frac{I_{out}}{\tilde{N}}~$,
where $\tilde{N}=\sum_{\alpha} Tr[\rho_{b} C(\alpha) \rho_{a}C^{\dagger}(\alpha)]~$.
Here, $\rho_{M}=|\phi \rangle_{M} \langle \phi|~$,
$\rho_{in \otimes out}=|\Psi \rangle_{in \otimes out} \langle \Psi|~$,
and $\tilde{\rho}_{M \otimes in}=|\tilde{\Psi} \rangle_{M \otimes in} \langle \tilde{\Psi}|~$.
Then, the conditional probability becomes

\begin{eqnarray}
P(\tilde{c}(\alpha)|\tilde{a},\tilde{b}) & = & \frac{Tr[\rho_{b} C(\alpha) \rho_{a} C^{\dagger}(\alpha)]}{\tilde{N}} \nonumber \\
& = & \frac{Tr[\tilde{\rho}_{M \otimes in} \otimes I_{out} C_{out}(\alpha) \rho_{M} \otimes \rho_{in \otimes out} C^{\dagger}_{out}(\alpha)]}{\tilde{N}^{2}} \nonumber \\
& = & \frac{Tr[C_{out} \tilde{\rho}_{M \otimes in} (\rho_{M} \otimes \rho_{in \otimes out}) C^{\dagger}_{out}]}{\tilde{N}^{2}} \nonumber \\
& = & Tr[C_{out} |\tilde{\phi} \rangle \langle \tilde{\phi}| C^{\dagger}_{out}],
\end{eqnarray}

%\end{document}
where
\begin{equation}
|\tilde{\phi} \rangle=\frac{1}{\tilde{N}}{}_{M \otimes in} \langle \tilde{\Psi}|(|\phi \rangle_{M} \otimes |\Psi \rangle_{in \otimes out}).
\end{equation}
%\end{document}
In equation (3), the dual state ${}_{M \otimes in} \langle \tilde{\Psi}|~$
can be regarded as a final state boundary condition at the future null infinity.
In this work, we investigate the thermodynamic properties of the outgoing Hawking radiation for a Schwarzschild black hole when the wave function of a black hole internal state is governed by the unitary evolution. We found that the Hawking radiation is suppressed when the final wave function is described by a coherent state. The condition for the black hole creation with coherent state as its internal state is investigated. The Hawking radiation of the Schwarzschild black hole is described by the Bogoliubov transformed vacuum for the Kruskal-Schwarzschild spacetime associated with a Schwarzschild black hole, which yield infinite dimensional two-mode squeezed state belonging to   $H_{in}\otimes H_{out}$ where $H_{in}~$ and $H_{out}~$ denote Hilbert spaces which contain quantum states localized inside and outside the event horizon, respectively \citep{Ahn1}.
The Hawking radiation
$|\Phi\rangle_{in\otimes out}$   belonging to $H_{in}\otimes
H_{out}~$ is given by \citep{Ahn1, Ahn2}

\begin{equation}
|\Phi\rangle_{in\otimes out}=\frac{1}{\cosh r_{\omega}}\sum_{n}e^{-4 \pi M\omega n}| l \rangle_{in}\otimes| l \rangle_{out}~,
\end{equation}
where $\{|{n}\rangle_{in}\}$  and  $\{|{n}\rangle_{out}\}$   are
orthonormal bases for  $H_{in}$ and $H_{out}$ , respectively, $M$ is the mass of the black hole, $\omega~$ is the positive frequency of the normal mode.
%\end{document}
%2010.10.21
The final wave function is given by \citep{Ahn1,Ahn2}

%equation (5)
\begin{equation}
{}_{M\otimes in}\langle\Psi|=\frac{1}{\cosh r_{\omega}} \sum_{n}e^{-4 \pi M \omega n}({}_{M}\langle {n} | S)\otimes {}_{in} \langle {n} |,
\end {equation}
where $S$ is a general unitary transformation which governs the evolution of the black hole internal state. In this model, we prescribe that the initial matter state $|\psi \rangle_{M} \in H_{M}~$, a pure state that will form the black hole, evolves into a state in $H_{M}\otimes H_{in} \otimes H_{out}~$, which is given by  $|\Psi_{0}\rangle_{M \otimes in \otimes out}=|\psi \rangle_{M} \otimes |\Phi \rangle_{in \otimes out}~$. The transformation from the quantum state of collapsing matter into the state of outgoing Hawking radiation when the black hole evaporates is given by the final state projection \citep{HM, Ahn1}:

%equation (6)
\begin{eqnarray}
|\phi \rangle_{out} & = & {}_{M \otimes in} \langle \Psi | \Psi_{0} \rangle_{M \otimes in \otimes} \nonumber \\
& = & \frac{1}{\cosh ^{2} r_{\omega}} \sum_{n,m} e^{-4 \pi M \omega (n+m)} {}_{M} \langle {m}|S| \psi \rangle_{M} {}_{in} \langle{m} | n \rangle_{in} \otimes |n \rangle_{out} \nonumber \\
& = & \frac{1}{\cosh ^{2} r_{\omega}} \sum_{n} e^{-8 \pi M \omega n} {}_{M} \langle{n}|S\psi \rangle_{M} |n \rangle_{out}.
\end{eqnarray}
%\end{document}
In weakly coupled string theory, the validity of semi-classical black hole description ( in which gravitation field is regarded as classical ) given above requires that the horizon size should be larger than the string length \citep{Giddings1,Horowitz}. In typical models where the string and Planck scales are not widely separated the above condition on the validity of a black hole description of a generic massive state produced in LHC is not significantly modified \citep{Giddings1}. Then the normalized outgoing state is given by

%2010.10.22.10:20AM

\begin{equation}
|\tilde{\phi} \rangle_{out}=\frac{1}{\sqrt{Z(\beta,\omega)}} ( \sum_{n}e^{-8 \pi M \omega n}|n \rangle_{out} {}_{M} \langle{n} |\psi \rangle_{M} ).
\end{equation}
%\end{document}	 				
The normalization factor $Z(\beta,\omega)~$ is defined by
\begin{equation}
Z(\beta,\omega)=\cosh ^{2} r_{\omega}\sqrt{{}_{out} \langle \phi | \phi \rangle_{out}}=\sum_{n}e^{-\beta \omega n} \vert {}_{M} \langle n |S|\psi \rangle_{M} \vert ^{2},
\end{equation}
%\end{document}	
where $\beta=16 \pi M~$.

For an arbitrary quantum mechanical operator $\hat{A}~$, we obtain

\begin{equation}
{}_{out} \langle \tilde{\phi} | \hat{A}| \tilde{\phi} \rangle_{out}=\sum_{n} A_{n} p_{n}(\beta,\omega).
\end{equation}
%\end{document}
Here the probability $p_{n}(\beta,\omega)~$ of finding the outgoing Hawking radiation in the normal mode  $n(\beta,\omega)~$ is defined by
%\end{document}

\begin{equation}
p_{n}(\beta,\omega)=\frac{1}{Z(\beta,\omega)} e^{- \beta \omega n}\vert {}_{M} \langle n | S | \psi \rangle_{M} \vert^{2}
\end{equation}
 %\end{document}
and $A_{n}~$  is an eigenvalue of $\hat{A}~$.
%\end{document}
%2010.10.22.11:45

The calculation of the average energy of the Hawking radiation is straight forward and is given by

\begin{eqnarray}
\langle E \rangle & = & \sum_{n} n \omega p_{n}(\beta,\omega) \nonumber \\
& = & \frac{1}{Z(\beta,\omega)} \sum_{n}n \omega e^{-\beta \omega n} \vert \langle n | S | \psi \rangle \vert ^{2} \nonumber \\
& = & -\frac{\partial}{\partial \beta} \log Z(\beta,\omega).
\end{eqnarray}
				 This is a well known thermodynamic relation of the average energy and the partition function when $Z(\beta,\omega)~$ is a partition function \citep{Page}.

%\end{document}

The number of expected emitting Hawking particle is given by

\begin{eqnarray}
\langle N \rangle & = & \sum_{n} n p_{n}(\beta,\omega) \nonumber \\
& = & \frac{\sum_{n}n e^{-\beta \omega n} \vert \langle n | S | \psi \rangle \vert ^{2}}{\sum_{n} e^{-\beta \omega n} \vert \langle n | S | \psi \rangle \vert ^{2}} \nonumber \\
& = & -\frac{\partial}{\beta \partial \omega} \log Z(\beta,\omega),
\end{eqnarray}
%\end{document}	 						
and the von Neumann entropy of the thermal radiation is described by
\begin{equation}
\delta S_{rad}=\sum_{n} \frac{e^{-\beta \omega n}}{Z(\beta,\omega)} \vert \langle n|S| \psi \rangle \vert^{2} ( \beta \omega n -2 \log \vert \langle n |S| \psi \rangle \vert + \log Z(\beta,\omega) ).
\end{equation}
%\end{document}
%2010.10.22.14:20
When the unitary operator $S$ is a random unitary, $\vert \langle n|S| \psi \rangle \vert^{2} \approx constant~$ \citep{Lloyd} and $Z_{Hawk}(\beta,\omega) \approx \sum_{n}e^{-\beta \omega n}~$, then we retrieve Hawking's original results \citep{Hawk2} with $\beta$  modified by a factor of two

\begin{equation}
\langle N \rangle_{Hawk}=-\frac{\partial}{\beta\partial \omega}Z(\beta,\omega)=\frac{1}{e^{\beta \omega}-1},
\end{equation}

%\end{document}
and
\begin{equation}
\delta S_{rad}=( \langle N \rangle_{Hawk}+1 ) \log ( \langle N \rangle_{Hawk}+1 ) - \langle N \rangle_{Hawk} \log \langle N \rangle_{Hawk} \geq 0.
\end{equation}
%\end{document}
Now we consider another extreme case in which the matter state $|\psi\rangle_{M}~$ which was built on the initial asymptotic vacuum state  $|0 \rangle_{M}~$ at the far past infinity $J^{-}~$ and evolves into a coherent state by a unitary generator $S(\alpha)~$. In this case, the black hole final state can be calculated analytically.
The operator which produces a coherence state $S(\alpha)=\exp (\alpha a^{\dag}-\alpha^{\ast}a)~$  where $a$  and $a^{\dag}$ are an annihilation operator and complex parameter, respectively, $S(\alpha)~$ is a unitary operator such that \citep{Barnett}

\begin{equation}
\langle n |S(\alpha)|0 \rangle = \exp(-\frac{|\alpha|^{2}}{2})\frac{\alpha^{n}}{\sqrt{n!}}.
\end{equation}
%\end{document}
Then, we obtain

\begin{eqnarray}
S(\alpha)|\psi \rangle_{M}=|\alpha \rangle_{M}=e^{-\frac{|\alpha|^{2}}{2}}\sum_{n}\frac{\alpha^{n}}{\sqrt{n!}} |n \rangle_{M},\nonumber \\
{}_{M} \langle n |S(\alpha) |\psi \rangle_{M}=\exp(-\frac{|\alpha|^{2}}{2})\frac{\alpha^{n}}{\sqrt{n!}}, \nonumber \\
Z_{CH}(\beta,\omega)=e^{-|\alpha|^{2}}\sum_{n}\frac{1}{n!} ( | \alpha |^{2}e^{-\beta \omega} )^{n}.
\end{eqnarray}
%\end{document}	 				

Then from equations (9) and (13), the number of expected emitted Hawking particle is given by

\begin{equation}
\langle N \rangle_{CH}=-\frac{\partial}{\beta \partial \omega}\log Z_{CH}(\beta,\omega)=\vert \alpha \vert^{2}e^{-\beta \omega}
\end{equation}
%\end{document}	 					
when the black hole matter state is described by the coherent state. The ratio of the Hawking radiation for these two extreme cases is given by

\begin{equation}
\langle N \rangle_{CH} / \langle N \rangle_{Hawk} = \vert \alpha \vert^{2}(1-e^{-\beta \omega}).
\end{equation}
%\end{document}
%2010.10.22 15:45
Above result suggests the ratio $\langle N \rangle_{CH} / \langle N \rangle_{Hawk}~$ becomes very small if the alpha parameter $|\alpha|\ll 1~$ or $\beta \omega \to 0~$. The latter would be the case for the black holes with mass in the TeV range ($M \approx 10^{-24} kg~$) produced in the LHC and the Hawking radiation up to the gamma ray spectrum ($10^{20} Hz~$).
%\end{document}
We define quadrature operator $\hat{x}_{\lambda}~$ by \citep{Barnett}
%\end{document}
\begin{equation}
\hat{x_{\lambda}}=\frac{1}{\sqrt{2}}( ae^{-i \lambda}+a^{\dag}e^{i \lambda} ).
\end{equation}
%\end{document}
%\end{document}

If we denote the eigenvector and eigenvalue of the quadrature operator by  $|x_{\lambda} \rangle~$ and $x_{\lambda}~$, respectively, then the basis state $|n \rangle~$ can be represented by

\begin{equation}
\langle x_{\lambda}|n \rangle = \pi^{-1/4}( 2^{n}n! )^{-1/2}e^{-x_{\lambda}^{2}-in\lambda}H_{n}(x_{\lambda}),
\end{equation}
%\end{document}
where $H_{n}~$ is a Hermite polynomial of order $n$ \citep{Barnett}. It is clear that $|x_{\lambda} \rangle~$ is equivalent to the familiar position representation of the state of a harmonic oscillator and $x_{\lambda}~$ can be used as an internal coordinate when the black hole state is given by a coherent state. The maximum value of $x_{\lambda}~$ can be taken as the radius of the black hole horizon. The quadrature eigenvalue  $x_{\lambda}~$ is related to the alpha parameter $\alpha$ by \citep{Barnett}

\begin{equation}
\langle x_{\lambda}^{2} \rangle = \langle 0| S^{\dag}(\alpha)\hat{x}^{2}_{\lambda}S(\alpha)|0 \rangle \approx |\alpha|^{2}.
\end{equation}
%\end{document}
Thus one can interpret $\langle x_{\lambda}^{2} \rangle~$ as a measure of wave function spread in the black hole.

For the TeV black hole generated with mass $M$ one can assume $\sqrt{\langle x_{\lambda}^{2} \rangle} \approx R_{H}~$ where the Schwarzschild radius $R_{H}~$ of $D=4+d~$  dimensions ($d=5~$ ) is \citep{Koch}

\begin{equation}
R_{H}=\ell_{p}|\frac{M_{P}}{M_{D}} ( \frac{M}{M_{D}}) \approx 2 \times 10^{-19} m
\end{equation}
%\end{document}
for the higher dimensional Planck mass  $M_{D} \approx 1 TeV~$ for the case of $M \approx M_{D}~$ where  $M_{P}~$ is the Planck mass.
%2010.10.23.01

For the black hole of Planck mass scale, the lifetime of a black hole on mass $M$ is given by \citep{Giddings1, Giddings3}

\begin{equation}
\tau_{Hawk}=C ( \frac{M}{M_{D}} ) ^{\frac{2D-3}{D-3}} \frac{1}{M},
\end{equation}
%\end{document}
where  C is a numerical constant.

For the TeV scale black hole created in the LHC, the relevant time scale for the black hole decay would be

\begin{equation}
\tau_{Hawk} \approx 1/M_{D} \approx 10^{-27} sec
\end{equation}
%\end{document}
for normal evaporation \citep{Giddings1}.

Since the emission of Hawking radiation would be suppressed by the factor  $\langle N \rangle_{CH} / \langle N \rangle_{Hawk}~$ when the black hole final wave function is described by the coherent state, the corresponding lifetime of the black hole would be given by
\begin{eqnarray}
\tau_{CH} & = & \langle N \rangle_{Hawk} / \langle N \rangle_{CH} \tau_{Hawk} \nonumber \\
& \approx & C \ell_{P}^{2} \frac{M_{P}^{2}M}{M_{D}^{4}}(\frac{M}{M_{D}})^{\frac{sD-4}{D-3}}.
\end{eqnarray}
%\end{document}
For the Hawking radiation with energy far greater than the gamma ray, say,  $10^{24}Hz~$, the suppression factor $\langle N \rangle_{CH} / \langle N \rangle_{Hawk}~$ is given by

\begin{equation}
\langle N \rangle_{CH} / \langle N \rangle_{Hawk} =| \alpha |^{2}(1-e^{-\beta \omega}) \approx 2.53 \times 10^{-38}
\end{equation}
%\end{document}
and the corresponding lifetime of evaporation $\tau_{CH}~$ is approximated by

\begin{equation}
\tau_{CH} = \langle N \rangle_{Hawk} / \langle N \rangle_{CH} / M_{D} \approx 4 \times 10^{11} sec \approx 10^{3} year
\end{equation}
%\end{document}
when a black hole internal wave function is described by the coherent state.

We also consider the case when the mass of the black hole is substantially greater than TeV to check the stability of the coherent state. For this we consider the case when $M \approx 1000 TeV (M \approx 1.8 \times 10^{-18} kg)~$. In this case, the D-dimensional Schwarzschild radius $R_{H}~$ is given by

\begin{equation}
R_{H} = \ell_{P} \frac{M_{P}}{M_{D}} ( \frac{M}{M_{D}} ) \approx 2 \times 10^{-16} m,
\end{equation}
%\end{document}
the the suppression factor $\langle N \rangle_{CH} / \langle N \rangle_{Hawk}~$  by

\begin{equation}
\langle N \rangle_{CH} / \langle N \rangle_{Hawk} =| \alpha |^{2}(1-e^{-\beta \omega}) \approx \langle x^{2}_{\lambda} \rangle (1-e^{-\beta \omega})\approx 4 \times 10^{-32}.
\end{equation}
%\end{document}
and the Hawking life time is given by

\begin{equation}
\tau_{Hawk}=C ( \frac{M}{M_{D}} ) ^{\frac{2D-3}{D-3}} \frac{1}{M} \approx 10^{-23} sec.
\end{equation}
%\end{document}
As a result, corresponding lifetime of black hole having coherent state is
\begin{equation}
\tau_{CH} = \langle N \rangle_{Hawk} / \langle N \rangle_{CH} \tau_{Hawk} \approx 2.5 \times 10^{8} sec \approx 8 year.
\end{equation}
%\end{document}

If the present model is a valid description of the black hole final quantum states, then our results demonstrate that the black hole evaporation is strongly dependent upon the black hole internal quantum states. Especially, when the internal matter state is represented by the coherent state, the black holes created in the LHC may have substantially long lifetimes.

\acknowledgements{This work was supported by the University of Seoul through the research grant of 2010 Seoul Metropolitan government.}

\end{document}